\documentclass[10pt,twocolumn,letterpaper]{article}

\usepackage{cvpr}
\usepackage{balance}  
\usepackage{color}    
\usepackage{url} 
\usepackage{amssymb}
\usepackage{amsmath}
\usepackage{caption}
\usepackage{listings}
\usepackage{paralist} 
\usepackage{llvm/lang} 
\usepackage{llvm/tblgen} 
\usepackage{nasm/style} 
\lstdefinestyle{mlir-code}{
  basicstyle=\fontsize{12}{12}\ttfamily,
  language=llvm, style=nasm,
  numbers=left, stepnumber=1, xleftmargin=2.5em,
  float=t,
  breaklines=true, breakatwhitespace=true,
}

\lstdefinestyle{mlir-code-nasm}{
  basicstyle=\fontsize{12}{12}\ttfamily,
  language=llvm, style=nasm,
  numbers=left, stepnumber=1, xleftmargin=1em,
  aboveskip=5pt, belowskip=5pt,
  breaklines=true, breakatwhitespace=true,
  frame=none, 
}

\lstdefinestyle{python-code}{
  basicstyle=\fontsize{12}{12}\ttfamily,
  language=python, style=nasm,
  numbers=left, stepnumber=1, xleftmargin=1em,
  aboveskip=5pt, belowskip=5pt,
  breaklines=true, breakatwhitespace=true,
  frame=none, 
}

\lstdefinestyle{c++-code}{
  basicstyle=\fontsize{12}{12}\ttfamily,
  language=c++, style=nasm,
  numbers=left, stepnumber=1, xleftmargin=1em,
  aboveskip=5pt, belowskip=5pt,
  breaklines=true, breakatwhitespace=true,
  frame=none, 
}

\lstdefinestyle{tblgen-code}{
  basicstyle=\fontsize{12}{12}\ttfamily,
  language=tblgen, style=nasm,
  numbers=left, stepnumber=1, xleftmargin=1em,
  aboveskip=5pt, belowskip=5pt,
  breaklines=true, breakatwhitespace=true,
  frame=none, 
}

\newcommand{\tensor}[1]{{\sf tensor}$\langle$#1$\rangle$}
\newcommand{\diop}[2]{{\sf #1}.{\sf #2}}  

\usepackage{booktabs}   
\cvprfinalcopy
\usepackage{graphicx}

\begin{document}

\title{TPU-MLIR: A Compiler For TPU Using MLIR}

\author{Pengchao Hu\qquad Man Lu\qquad Lei Wang\qquad Guoyue Jiang\\
{\tt\small \{pengchao.hu,man.lu,lei.wang,guoyue.jiang\}@sophgo.com}\\
Sophgo Inc.
}

\maketitle

\begin{abstract}
  Multi-level intermediate representations (MLIR) show great promise for reducing the cost of building domain-specific compilers by providing a reusable and extensible compiler infrastructure.
This work presents TPU-MLIR, an end-to-end compiler based on MLIR that deploys pre-trained neural network (NN) models to a custom ASIC called a Tensor Processing Unit (TPU).
TPU-MLIR defines two new dialects to implement its functionality:
\begin{inparaenum}
\item a Tensor operation (TOP) dialect that encodes the deep learning graph semantics and independent of the deep learning framework and
\item a TPU kernel dialect to provide a standard kernel computation on TPU.
\end{inparaenum}
A NN model is translated to the TOP dialect and then lowered to the TPU dialect for different TPUs according to the chip's configuration.
We demonstrate how to use the MLIR pass pipeline to organize and perform optimization on TPU to generate machine code.
The paper also presents a verification procedure to ensure the correctness of each transform stage.

\end{abstract}

\section{Introduction}\label{sec:intro}
The development of deep learning (DL) has profoundly impacted various scientific fields, including speech recognition, computer vision, and natural language processing.
In order to facilitate the process of training deep learning models, industry and academia have developed many frameworks, such as Caffe, Tensorflow, Pytorch, Mxnet, and PaddlePaddle, which boost deep learning in many areas.
However, each framework has its proprietary graph representation, which brings lots of work for deploying as we need to support many DL model formats.

At the same time, matrix multiplication and high dimensional tensor convolution are the heavy computation in DL, which evoke the passion of chip architects to design customized DL accelerators to achieve high performance at low energy.
Although GPU is still the leading hardware in training DL models and all the DL frameworks have contributed much work to support this general-purpose hardware, GPU is not the perfect piece in the inference domain of DL.
GPU is for gaming, graph rendering, scientific computation, and much more, not tailored for DL only.
Thus, many DL accelerators, such as Google TPU, Apple Bonic, Graphcore IPU, and SOPHGO TPU, are more energy efficient than GPU and benefit many of these emerging DL applications.

In addition, the DL community has resorted to domain-specific compilers for rescue to address the drawback of DL libraries and alleviate the burden of manually optimizing the DL models on each DL hardware.
The DL compilers take the model described in the DL frameworks as inputs and generate efficient code for various DL hardware as outputs.
The transformation between a model definition and specific code implementation is highly optimized, considering the model specification and hardware architecture.
Several popular DL compilers, such as TVM, Tensor Comprehension, and XLA, have been proposed by industry and academia.
Specifically, they incorporate DL-oriented optimizations such as layer and operator fusion, which enables highly efficient code generation.

Herein, We provide TPU-MLIR, an open-source DL compiler for TPU. In particular, we chose Open Neural Network Exchange (ONNX)\cite{bai2019} as a DL format to represent our compiler's input model and use Multi-level Intermediate Representation (MLIR)~\cite{lattner2021mlir}, a modern open-source compiler infrastructure for multi-level intermediate representation, to design TPU-MLIR\footnote{https://github.com/sophgo/tpu-mlir} compiler.

In this work, we will introduce our compiler by

\begin{itemize}
\item presenting the overall design and architecture of the compiler,
\item introducing two new dialects: TOP dialect to encode the deep learning graph semantics independent of the deep learning framework and TPU dialect to provide a common lowering point for all TOP dialect operations but device-dependent,
\item detailing each compile stage, such as converting NN models to Top dialect as device independent and then converting TOP to TPU for various chips and types,
\item defining WeightOp for weight operation and store weight data in the NumPy npz file, and
\item providing {\sf InferenceInterface} for TOP and TPU to ensure correct conversions.
\end{itemize}

We organize the remainder of the paper as follows.
In Sec.~\ref{sec:background}, we briefly discuss MLIR, ONNX, on which our compiler is based, and the calibration processing, which tailors computation for TPU.
Sec.~\ref{sec:tpu-mlir}, we introduce our compiler's design principle and architecture and discuss TOP and TPU dialects.
We also discuss using inference to ensure correctness in each conversion stage.
Finally, we conclude our paper and discuss future work in Sec.~\ref{sec:conclusion}.
%


\section{Background}\label{sec:background}
\subsection{MLIR}
The MLIR, with much reusable and extensible, is a novel approach for constructing new domain-specific compilers.
An open ecosystem is the most significant difference from LLVM.
MLIR standardizes the Static Single Assignment (SSA)-based IR data structures allowing one to express a range of concepts as first-class operations.
Operations can represent many different levels of abstraction and computations, from dataflow graphs to target-specific instructions and even hardware circuitry.
They take and produce zero or more values, called operands and results, respectively. A value represents data at runtime and is associated with a type known at compile-time, whereas types model compile-time information about values.
Complementary to this, attributes contain compile-time information to operations. Operations, Attributes, and type systems are open and extensible.
The custom types, operations, and attributes are logically grouped into dialects. A dialect is one of the most fundamental aspects of MLIR that enables the infrastructure to implement a stack of reusable abstractions.
Each abstraction encodes and preserves transformation validity preconditions directly in its IR, reducing the complexity and cost of analysis passes.
The MLIR IR has a recursive structure where operations contain a list of regions, and regions contain a list of blocks, which in turn, contain a list of operations.

In particular, MLIR features operation, attribute and type interfaces providing a generic way of interacting with the IR.
Interfaces allow transformations and analyses to work with abstract properties rather than fixed lists of supported concepts.
Interfaces can be implemented separately from operations and mixed in using MLIR's registration mechanism, thus fully separating IR concepts from transformations.
Furthermore, transformations can be written as compositions of orthogonal localized "match and rewrite" primitives.
These are often decomposed further into rewriting rules when applied within a dialect and lowering rules when converting from a higher-level dialect to a lower-level dialect.
Throughout the compilation, separate dialects can co-exist to form a hybrid program representation.
The ability to progressively lower dialects to the target hardware during the compilation process has made MLIR an excellent compiler infrastructure for domain-specific languages.

This article relies on several MLIR dialects and types, briefly described below.

\subsubsection{Ranked Tensor Type}
Values with tensor type represent aggregate N-dimensional homogeneous data indicated by element type and a fixed rank with a list of dimensions\footnote{https://mlir.llvm.org/docs/Dialects/Builtin/\#rankedtensortype}.
Each dimension could be a static non-negative integer constant or be dynamically
determined (marked by {\sf ?}).

This abstracted runtime representation carries both the tensor data values and information about the tensor shape, but the compiler has not decided on its representation in memory. Tensor values are immutable and subject to {\sf def-use} SSA semantics\cite{vasilache2022composable}.
Operations on tensors are often free of side effects, and operations always
create new tensors with a value.
The textual format of the tensor is \tensor{$\sf d_1 x d_2x  \cdots x d_N x dtype$},
where $\sf{d}_1$, $\sf{d}_{2}$, ... $ \sf{d}_{\sf{N}}$ are integers or symbol
{\sf ?} representing the dimensions of a tensor, and {\sf dtype} is the type of the
elements in a tensor, e.g., F32 for float32.
A tensor can be unranked when its shapes are unknown. MLIR uses \tensor{$\sf *{x}dtype$} to represent unranked tensor types.

\subsubsection{Quantization Dialect}
Quantization dialect\footnote{https://mlir.llvm.org/docs/Dialects/QuantDialect} provides a family of quantized types and type-conversion operations.
The "quantization" refers to the conversion of floating-point computations to
corresponding variants expressed in integer math for inference, as has been
supported by low-bit depth inference engines such as various accelerator
hardware and many DSPs.
There are three types defined in quantization dialect: UniformQuantizedType, UniformQuantizedPerAxisType, and CalibratedQuantizedType.
The UniformQuantizedType and  UniformQuantizedPerAxisType represent the mapping
between expressed values (e.g., a floating-point computer type) and storage
values (typically of an integral computer type), expressing the affine
transformations from uniformly spaced points to the real number line.
The relationship is: $\sf realValue = scale \times (quantizedValue - zeroPoint) $ and will be discussed in more detail in Section~\ref{sec:quant}.
Where CalibratedQuantizedType holds the range from the given min and max value of the histogram data of the tensor, used for recording the statistics information of the tensor.
The UniformQuantizedPerAxisType applies affine transformation individually to each index along a specific axis of a tensor type.
However, UniformQuantizedType applies the affine transformation to every value within the target type.
The type-conversion defined in quantization dialect provides three operations for converting bet%
ween types based on a QuantizedType and its expressed and storage sub-types.
Those operations are: \diop{quant}{qcast} converting from an expressed type to QuantizedType, \diop{quant}{dcast} converting from a QuantizedType to its expressed type, and \diop{quant}{scast} converting between a QuantizedType and its storage type.

\subsection{ONNX}
ONNX is an open-source framework-independent format widely used for exchanging computation graph models, including deep learning and traditional machine learning.
It was accepted as a graduate project in Linux Foundation AI and maintained by open-source communities.
ONNX defines an extensible computation graph model, operators, and standard data types for deep learning and provides a set of specifications to convert a model to a basic ONNX format and another to get the model back from this ONNX form.
It is an ideal tool for framework interoperability, especially when deploying a model to specific hardware\cite{jin2020compiling}.

ONNX reduces the friction of moving trained DL models among AI frameworks and platforms.
ONNX uses the Protocol Buffers language for its syntax and provides rich documents and tools to formalize each operation's semantics and verify its correctness.

\subsection{Quantization}\label{sec:quant}
Quantization is a promising technique to reduce deep learning models' memory footprint, inference latency, and power consumption, which replaces high-cost floating-point (always F32) computation with low-cost fixed-point numbers\cite{jacob2018quantization} (e.g., INT8/INT16) or float-point (e.g., BF16/F16).
Because most current DL models are heavily over-parameterized and robust to extreme discretization, there is much opportunity for reducing numeral precision without impacting the model's accuracy, bringing ample search space for tuning.
Although many quantization methods have emerged, there is not a single well-posed or well-conditioned problem being solved\cite{gholami2021survey}.
Instead, one is interested in some error metric (based on classification quality, data similarity, etc.).
to guide the quantization process.
However, due to the over-parameterization, it is possible to have a high error between a quantized and the original model while still attaining excellent generalization performance.
Finally, different layers in a Neural Net have a different impact on the loss function, which motivates a mixed-precision approach quantization.

\subsubsection{Uniform Quantization}
The quantization process is a function mapping from real values $r$ to some numeral values.
Quantization function such as
\begin{equation}
  \label{equ:quant}
  {\sf quant}(r) = {\sf round}(\frac{r}{s}) + {\sf zp}
\end{equation}
where $\text{quant}$ is the quantization operator, $r$ is a real-valued input
(activation or weight), $s$ is a float-point scaling factor, and $\sf zp$ is an
integer zero point, is known as uniform quantization, as the resulting quantized
values are evenly spaced.

\subsubsection{Symmetric and Asymmetric Quantization}
A crucial factor in uniform Quantization is choosing the scaling factor $s$ in Equation~\ref{equ:quant}.
This scaling factor, also known as resolution, divides a given range of real-values r into several partitions $s = \frac{\beta -  \alpha}{2^b - 1}$, where $[\alpha, \beta]$ denotes the clipping range that we are clipping the real-values with, and b is the quantization bit width\cite{jacob2018quantization}\cite{krishnamoorthi2018quantizing}.
Therefore, one should determine the clipping range $[\alpha, \beta]$ before generating the scaling factor.
If the clipping range of $\alpha$ equals $-\beta$, we get Symmetric Quantization, and on the contrary, we get asymmetric Quantization.
The asymmetric quantization method often results in a tighter clipping range than symmetric Quantization, which is especially important when the dynamic range of the tensor is imbalanced, e.g., the result of ReLU always has non-negative values.

\subsubsection{Calibration}
The process of choosing the clipping range is called ``calibration.''
One popular method for pre-calculation is to run a series of inferences on some sample data and then get the distribution of each tensor in the graph.
Using the min/max of the signal for both symmetric and asymmetric Quantization is typical in most cases.
However, this approach is susceptible to outlier data in the activations, which could unnecessarily increase the range and reduce quantization resolution.
One approach to address this is using percentile or selecting $\alpha$ and $\beta$ to minimize KL divergence between the real and the quantized values\cite{nvidia8bit}\cite{wu2020integer}.
Besides, there are other metrics to find the best range, including minimizing Mean Squared Error (MSE)\cite{choukroun2019low}, entropy, and cosine similarity.


\section{Compiler}\label{sec:tpu-mlir}
This section introduces the compiler, TPU-MLIR, which creates two layers by the TOP and TPU dialects for converting NN models to executable files by various types and chips.
We discuss TPU-MLIR's overall architecture first.

\begin{figure}[t]
   \centering
   \includegraphics[scale=0.35]{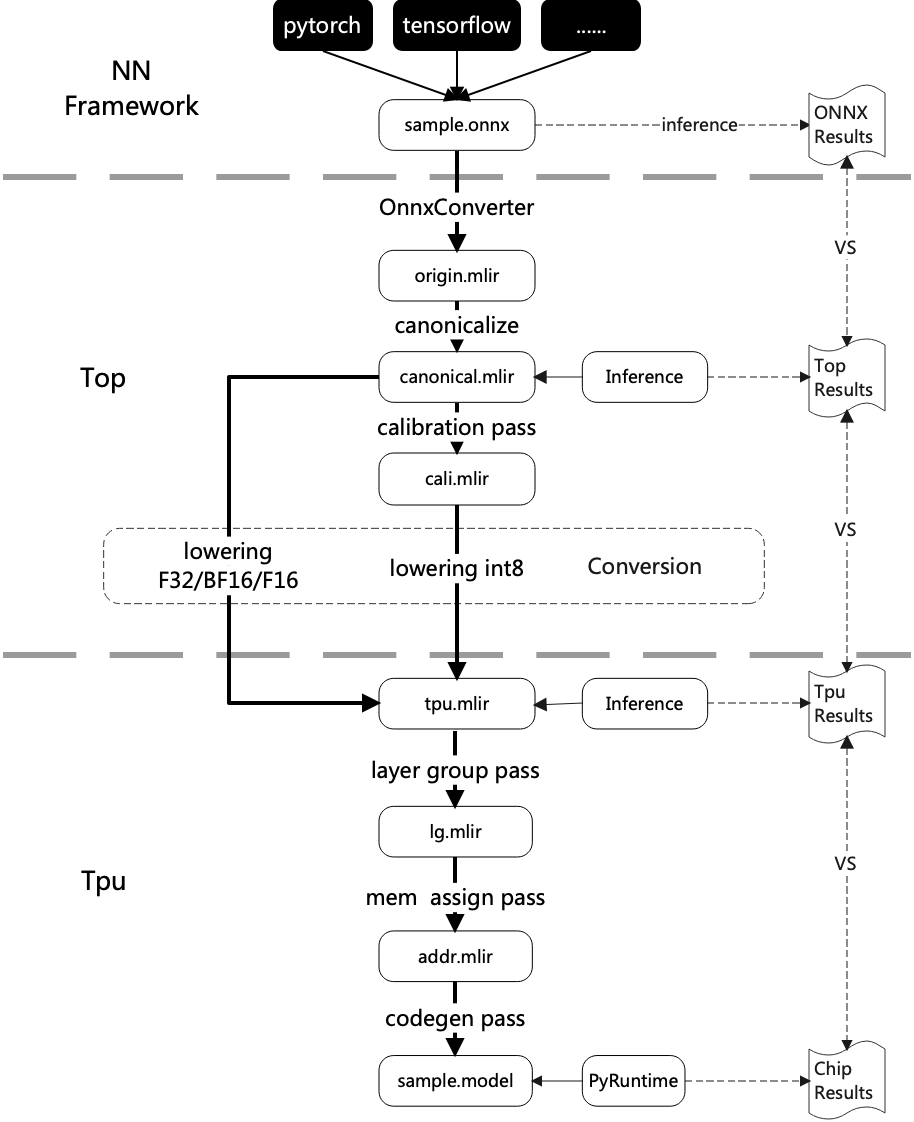}
   \caption{Architecture of tpu-mlir.}
   \label{fig:tpu-mlir}
\end{figure}

\subsection{Overview}

Figure~\ref{fig:tpu-mlir} shows the overall architecture of TPU-MLIR. We divide it into the NN Framework, Top, and Tpu.
\begingroup
\begin{enumerate}
\renewcommand\labelenumi{\theenumi)}
  \item {\bf NN Framework}: TPU-MLIR supports ONNX models directly. Other NN framework models, such as Pytorch, and Tensorflow, need to convert to ONNX modes.
  \item {\bf TOP}: refer to the TOP dialect as the top abstraction level representing NN models in the MLIR language. It is device independent.
  \item {\bf TPU}: refer to the TPU dialect, which is the TPU abstraction level and represents TPU operations. It is device dependent.
\end{enumerate}
\endgroup

We first convert a NN model to TOP abstraction with TOP dialect and built-in dialect defined in MLIR, which we call TOP mlir file, by python script, i.e., OnnxConverter in figure~\ref{fig:tpu-mlir}.
Then we lower the top mlir file to TPU abstraction with TPU dialect and built-in dialect defined in MLIR, which we call tpu mlir through some passes, such as canonicalization pass and calibration pass.
At last, we convert tpu mlir to tpu models by some passes, such as layer group pass and memory assign pass.
These passes will be discussed in the later section.

\subsection{Module}

We introduce our module definition by a simple mlir file showed Listing~\ref{lst:sample-model}:

\begin{lstlisting}[style=mlir-code, caption={Simple convolution computation represented by TPU dialect.}, label={lst:sample-model}, frame=lines, float=*t]
#loc0 = loc(unknown)
module attributes {module.name = "Sample", module.weight_file = "conv2d_weight.npz"} {
  func.func @main(%arg0: tensor<1x16x100x100xf32> loc(unknown)) -> tensor<1x32x100x100xf32> {
    %0 = "top.None"() : () -> none loc(#loc0)
    %1 = "top.Input"(%arg0) : (tensor<1x16x100x100xf32>) -> tensor<1x16x100x100xf32> loc(#loc1)
    %2 = "top.Weight"() : () -> tensor<32x16x3x3xf32> loc(#loc2)
    %3 = "top.Weight"() : () -> tensor<32xf32> loc(#loc3)
    %4 = "top.Conv"(%1, %2, %3) {dilations = [1, 1], do_relu = false, group = 1 : i64, kernel_shape = [3, 3], pads = [1, 1, 1, 1], strides = [1, 1]} : (tensor<1x16x100x100xf32>, tensor<32x16x3x3xf32>, tensor<32xf32>) -> tensor<1x32x100x100xf32> loc(#loc4)
    return %4 : tensor<1x32x100x100xf32> loc(#loc0)
  } loc(#loc0)
} loc(#loc0)
#loc1 = loc("input")
#loc2 = loc("filter_conv1")
#loc3 = loc("bias_conv1")
#loc4 = loc("conv1")
\end{lstlisting}

Module has some attributes: \diop{module}{name} is related to the NN model name; \diop{module}{weight\_file} is a npz\footnote{https://numpy.org/neps/nep-0001-npy-format.html} file that stores weight data needed by operations.
We use location to express operation name.
For example, '{\%2 = ``top.Weight''()}' (Line~$6$ in Listing~\ref{lst:sample-model}) is a weight op, and location is ``filter\_conv1''. So the real weight data is stored in ``conv2d\_weight.npz'' file by name ``filter\_conv1''.

\subsection{Top Dialect}

TOP dialect is very similar to TOSA (Tensor Operator Set Architecture)\footnote{https://www.mlplatform.org/tosa} dialect in MLIR.
So why we don't use TOSA dialect? There are two reasons: the first is that we need to do inference for each operations, and may create some new features in the futrue; the second is that we need to keep extend capability to support various NN models.

TOP Dialect is defined as below:

\begin{lstlisting}[style=tblgen-code]
def Top_Dialect : Dialect {
  let name = "top";
  let summary = "A top dialect for the TPU_MLIR specification";
  let cppNamespace = "::tpu_mlir::top";
  let emitAccessorPrefix = kEmitAccessorPrefix_Raw;
  let extraClassDeclaration = [{...}];
}
\end{lstlisting}

In TOP Dialect,  TOP\_BaseOp and TOP\_Op define as:

\begin{lstlisting}[style=tblgen-code]
class Top_BaseOp<string mnemonic, list<Trait> traits = []> :
    Op<Top_Dialect, mnemonic, !listconcat(traits,[NoSideEffect])>;

class Top_Op<string mnemonic, list<Trait> traits = []> :
Top_BaseOp<mnemonic, !listconcat(traits,
[DeclareOpInterfaceMethods<InferenceInterface>,
DeclareOpInterfaceMethods<FlopsInterface>])>;
\end{lstlisting}

TOP\_Op has two interfaces: ``InferenceInterface'' and ``FlopsInterface''. ``InferenceInterface'' is used to do inference for operation, which would be introduced later. ``FlopsInterface'' is used to count FLOPs (floating point operations) of operation, as we are interested in the FLOPs of a NN model, also we use it to evaluate chip performance after running on the chip.

There are top operations defined based on TOP\_BaseOp or TOP\_Op. Here just using ConvOp and WeightOp for examples.

\subsubsection{top::ConvOp}

ConvOp is defined as below:

\begin{lstlisting}[style=tblgen-code]
def Top_ConvOp: Top_Op<"Conv"> {
  let summary = "Convolution operator";
  let arguments = (ins
    AnyTensor:$input,
    AnyTensor:$filter,
    AnyTensorOrNone:$bias,
    I64ArrayAttr:$kernel_shape,
    I64ArrayAttr:$strides,
    I64ArrayAttr:$pads, // top,left,bottom,right
    DefaultValuedAttr<I64Attr, "1">:$group,
    OptionalAttr<I64ArrayAttr>:$dilations,
    DefaultValuedAttr<BoolAttr, "false">:$do_relu,
    DefaultValuedAttr<F64Attr, "-1.0">:$relu_limit
  );
  let results = (outs AnyTensor:$output);
}
\end{lstlisting}

ConvOp represents conv operation of NN models, like Figure~\ref{fig:convOp}:

\begin{figure}[t]
   \centering
   \includegraphics[scale=0.4]{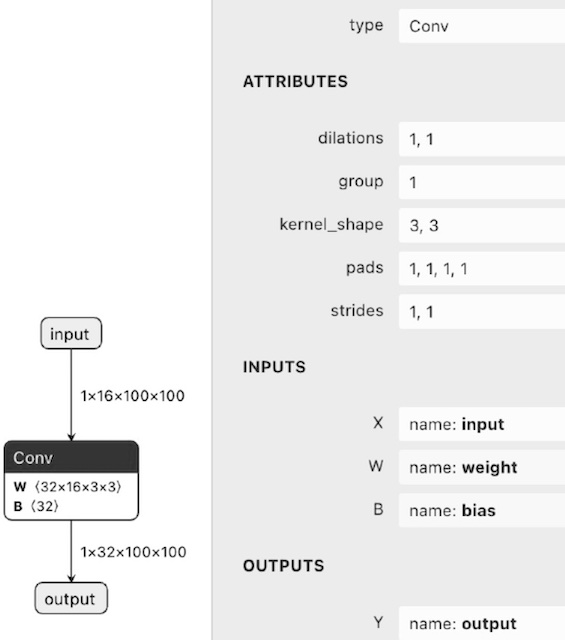}
   \caption{Convolution operation defined in ONNX.}
   \label{fig:convOp}
\end{figure}

and in mlir file experessed as below:

\begin{lstlisting}[style=mlir-code-nasm]
%4 = "top.Conv"(%1, %2, %3) {dilations = [1, 1], do_relu = false, group = 1 : i64, kernel_shape = [3, 3], pads = [1, 1, 1, 1], strides = [1, 1]} : (tensor<1x16x100x100xf32>, tensor<32x16x3x3xf32>, tensor<32xf32>) -> tensor<1x32x100x100xf32>
\end{lstlisting}

\subsubsection{top::WeightOp}

WeightOp is a special operation for weight datas. Defined as below:

\begin{lstlisting}[style=tblgen-code]
def Top_WeightOp : Top_BaseOp<"Weight"> {
  let summary = "load weight operator";
  let results = (outs AnyTensor:$output);
  let extraClassDeclaration = [{
  template<typename T>
  std::shared_ptr<std::vector<T>> read();
  template<typename T>
  static mlir::Value create(mlir::Operation * OwnerOp,
    llvm::StringRef suffix,
    const std::vector<T>& data,
    mlir::RankedTensorType& type);
  }];
}
\end{lstlisting}

WeightOp is corresponding to weight operation. Weight data is stored in ``module.weight\_file'', WeightOp can read data from weight file by {\bf read} method, or create new WeightOp by {\bf create} method.

\subsection{TPU Dialect}

TPU Dialect is defined as below:

\begin{lstlisting}[style=tblgen-code]
def Tpu_Dialect : Dialect {
  let name = "tpu";
  let summary = "A tpu dialect for the SOPHGO AI chips";
  let cppNamespace = "::tpu_mlir::tpu";
  let useDefaultAttributePrinterParser = 1;
  let emitAccessorPrefix = kEmitAccessorPrefix_Raw;
}
\end{lstlisting}

TPU dialect is for TPU chips, here we support SOPHGO AI chips first. It is used to generate chip command instruction sequences by tpu operations.

In TPU dialect,  TPU\_BaseOp and TPU\_Op define as:

\begin{lstlisting}[style=tblgen-code]
class Tpu_BaseOp<string mnemonic, list<Trait> traits = []> :
  Op<Tpu_Dialect, mnemonic,
     !listconcat(traits,[NoSideEffect, TpuTypeRestrict])> ;

class Tpu_Op<string mnemonic, list<Trait> traits = []> :
  Op<Tpu_Dialect, mnemonic, !listconcat(traits,
    [NoSideEffect, TpuTypeRestrict,
      DeclareOpInterfaceMethods<
        GlobalGenInterface>,
      DeclareOpInterfaceMethods<
        InferenceInterface>])>;
\end{lstlisting}

TPU\_Op has two interfaces, ``GlobalGenInterface'' and ``InferenceInterface''. ``GlobalGenInterface'' is used to generate chip command. ``InferenceInterface'' is used to do inference for tpu operations.

There are top operations defined based on TOP\_BaseOp or TOP\_Op. Here using tpu::ConvOp , and tpu::CastOp, and tpu::GroupOp for example.

\subsubsection{tpu::Conv2DOp}

Conv2DOp is defined as below:

\begin{lstlisting}[style=tblgen-code]
def Tpu_Conv2DOp: Tpu_Op<"Conv2D"> {
  let arguments = (ins
    AnyTensor:$input,
    AnyTensor:$filter,
    AnyTensorOrNone:$bias,
    I64ArrayAttr:$kernel_shape,
    I64ArrayAttr:$strides,
    I64ArrayAttr:$pads, // top,left,bottom,right
    DefaultValuedAttr<I64Attr, "1">:$group,
    OptionalAttr<I64ArrayAttr>:$dilations,
    DefaultValuedAttr<BoolAttr, "false">:$do_relu,
    DefaultValuedAttr<F64Attr, "-1.0">:$relu_limit,
    //new param
    OptionalAttr<I64ArrayAttr>:$multiplier,
    OptionalAttr<I64ArrayAttr>:$rshift,
    OptionalAttr<Tpu_LayerGroupAttr>:$group_info
  );

  let results = (outs AnyTensor:$output);
}
\end{lstlisting}

Compared to top::ConvOp, tpu::Conv2DOp has some new attributes: ``multiplier'', ``rshift'' and ``group\_info''. ``multiplier'' and ``rshift'' are used to do INT8 convolution after quantization, and not used if the convolution is float. ``group\_info'' is used for the layer group. We will discuss layer group later.

\subsubsection{tpu::CastOp}

\begin{lstlisting}[style=mlir-code, caption={\diop{top}{cast} operaiotn convertion a \diop{quant}{calibrated} type to \diop{quant}{uniform} type.}, label={lst:tpu-cast}, frame=lines, float=*t]
%1 = "tpu.Cast"(%0) : (tensor<1x16x100x100x!quant.calibrated<f32<-4.1780586242675781:4.4932479858398438>>>) -> tensor<1x16x100x100x!quant.uniform<i8:f32, 0.034005123961205579:-5>>
\end{lstlisting}

\begin{lstlisting}[style=mlir-code, caption={\diop{top}{cast} operaiotn convertion a \diop{quant}{uniform} type to float32 type.}, label={lst:tpu-quant}, frame=lines, float=*t]
%5 = "tpu.Cast"(%4) : (tensor<1x32x100x100x!quant.uniform<i8:f32, 0.43113517013250613:-2>>) -> tensor<1x32x100x100xf32>
\end{lstlisting}

CastOp is defined as below:

\begin{lstlisting}[style=tblgen-code]
def Tpu_CastOp:Tpu_Op<"Cast",
  [DeclareOpInterfaceMethods
    <LocalGenInterface>]> {
  let summary = "Cast operation";
  let description = [{}];
  let arguments = (ins
    AnyTensor:$input,
    OptionalAttr<Tpu_LayerGroupAttr>:$group_info
  );
  let results = (outs AnyTensor:$output);
}
\end{lstlisting}

CastOp is for transferring tensor type from one type to another. It can convert the F32 type to BF16\cite{wang2019bfloat16} type or F16 type, or INT8 type, and the other way around is also OK.

Specially, if input is F32 type and  output is quantization type, such as Listing~\ref{lst:tpu-cast}, then:

\begin{equation}
  \sf{output} = RoundToInt8( \frac{input_{f32}}{qscale} + zeroPoint)
\end{equation}

If input is quantization type and output is F32 type, such as
Listing~\ref{lst:tpu-quant}, then
\begin{equation}
  \sf output = (input_{i8} - zeroPoint) * qscale
\end{equation}.

\subsubsection{tpu::GroupOp}

GroupOp is defined as below:

\begin{lstlisting}[style=tblgen-code]
def Tpu_GroupOp:Tpu_BaseOp<"Group"> {
  let summary = "Group operation";
  let description = [{
    Make ops in one group to inferece by local mem
  }];
  let arguments = (ins
    Variadic<AnyTensor>:$inputs,
    I64Attr:$nsecs,
    I64Attr:$hsecs
  );
  let results = (outs Variadic<AnyTensor>:$outputs);
  let regions = (region SizedRegion<1>:$body);
}
\end{lstlisting}

GroupOp contains serial operations that can inference in tpu local memory. We will discuss it later.

\subsection{Conversion}

This section we discuss how to convert top ops to tpu ops.

We define ``ConvertTopToTpu'' pass like this:

\begin{lstlisting}[style=tblgen-code]
def ConvertTopToTpu : Pass<"convert-top-to-tpu", "ModuleOp"> {
  let summary = "Convert top-level Top Ops to Tpu Ops";
  let constructor = "tpu_mlir::createConvertTopToTpu()";
  let dependentDialects = ["tpu_mlir::top::TopDialect", "tpu_mlir::tpu::TpuDialect"];
  let options = [
    Option<"mode", "mode", "std::string", /*default=*/"",
           "default quantization mode: INT8/BF16/F16/F32">,
    Option<"chip", "chip", "std::string", /*default=*/"",
           "chip: cv183x/cv182x/bm1684/bm1684x">,
    Option<"isAsymmetric", "asymmetric", "bool", /*default=*/"false",
           "true for asymmetric quantization, or false for symmetric">,
  ];
}
\end{lstlisting}

There are there options: mode, chip and isAsymmetric.
\begingroup
\begin{enumerate}
\renewcommand\labelenumi{\theenumi)}
\item {\bf mode}: set quantization mode, e.g. INT8, BF16, F16 or F32. Types should be supported by the chip.
\item {\bf chip}: set chip name. TPU operations will act by this chip.
\item {\bf isAsymmetric}: if {\bf mode} is INT8, set true for asymmetric quantization; false for
symmetric quantization.
\end{enumerate}
\endgroup

\begin{figure}[t]
   \centering
   \includegraphics[scale=0.3]{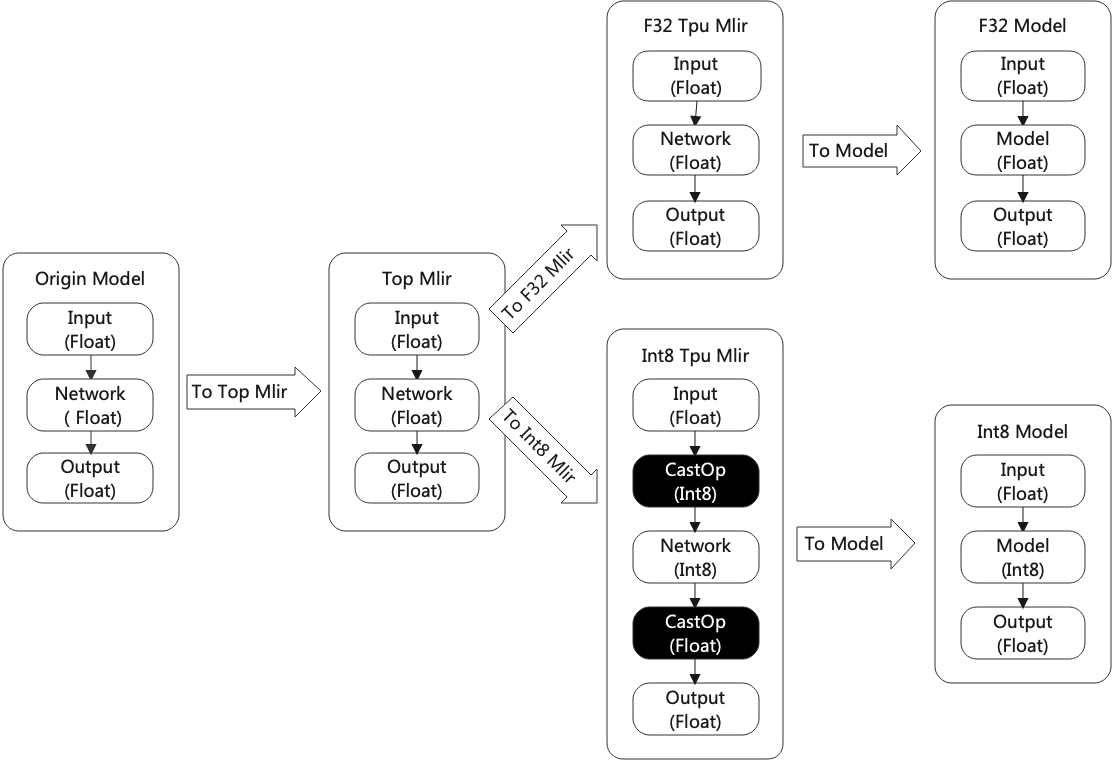}
   \caption{The neural network model conversion flow of TPU-MLIR.}
   \label{fig:flow}
\end{figure}

Typically, most attributes are the same after converting from TPU ops to TPU ops at float type (F32/BF16/F16). However, if the conversion is from float type to INT8 type, we should do PTQ (Post-training Quantization)\cite{jacob2018quantization} at TOP dialect and add some external quantization attributes to TPU ops. At the same time, weight data, inputs, and outputs will be quantized to INT8. In addition, inputs and outputs of a NN model need to insert CastOp if the convert type is not F32. The conversion flow chart shows in Figure~\ref{fig:flow}.

\subsection{Inference}

This section discusses why we need inferences and how to support inferences for TOP and TPU dialects.

\subsubsection{Why}

TOP dialect run inference and get inference results, which has three uses.
\begingroup
\begin{enumerate}
\renewcommand\labelenumi{\theenumi)}
\item It can be used to compare with original model results, to make sure NN model converts to TOP dialect correctly.
\item It can be used for calibration, which uses a few sampled inputs to run inference by top mlir file and get every intermediate result to stat proper min/max threshold used by Quantization.
\item It can be used to compare with the inference results tpu dialect to ensure tpu mlir is correct.
\end{enumerate}
\endgroup

TPU dialect runs inference and gets inference results, which would compare with top mlir results. If tpu mlir is in F32 mode, the results should be the same. If tpu mlir is BF16/F16 mode, the tpu results may have some loss but should still have a good cosine ($>$0.95) and euclidean ($>$0.85) similarity. If tpu mlir is INT8 mode, cosine similarity should be greater than 0.9, and euclidean similarity should be greater than 0.5, based on experience. If the cosine similarity and euclidean similarity are not satisfied, the conversion correction from top to tpu is not guaranteed.

Cosine similarity is defined as below:

\begin{lstlisting}[style=python-code]
def square_rooted(self, x):
  return sqrt(sum([a*a for a in x]))

def cosine_similarity(self, x, y):
  numerator = sum(a*b for a,b in zip(x,y))
  denominator = self.square_rooted(x)*self.square_rooted(y)
  return round(numerator/float(denominator),3)

cosine_similarity = cosine_similarity(x, y)
\end{lstlisting}

Euclidean similarity is defined as below:

\begin{lstlisting}[style=python-code]
def euclidean_distance(x, y):
  return sqrt(sum(pow(a-b,2) for a, b in zip(x, y)))
ed = euclidean_distance(x, y)
sr = square_rooted((x+y)/2)
euclidean_similary = 1 - ed/sr
\end{lstlisting}

At last, after being compiled, the model can deploy in the tpu device and check the result with tpu mlir to ensure codegen is correct. If not similar, there are some bugs in codegen.

\subsubsection{How}

The NN models will run on NN runtime. For example, ONNX models can run on ONNX runtime. TOP dialect and TPU dialect run inference by ``InferenceInterface'', which defines as below:

\begin{lstlisting}[style=c++-code]
struct InferenceParameter {
  std::vector<float *> inputs;
  std::vector<float *> outputs;
  void *handle = nullptr;
};
\end{lstlisting}

\begin{lstlisting}[style=tblgen-code]
def InferenceInterface : OpInterface<"InferenceInterface"> {
  let cppNamespace = "::tpu_mlir";
  let methods = [
      InterfaceMethod<
        /*desc=*/[{}],
        /*retType=*/"::mlir::LogicalResult",
        /*methodName=*/"inference",
        /*args=*/(ins "InferenceParameter&":$param)
      >
      InterfaceMethod<
        /*desc=*/[{}],
        /*retType=*/"::mlir::LogicalResult",
        /*methodName=*/"init",
        /*args=*/(ins "InferenceParameter&":$param)
      >,
      InterfaceMethod<
        /*desc=*/[{}],
        /*retType=*/"void",
        /*methodName=*/"deinit",
        /*args=*/(ins "InferenceParameter&":$param)
      >,
  ];
}
\end{lstlisting}

\begin{figure}[t]
   \centering
   \includegraphics[scale=0.4]{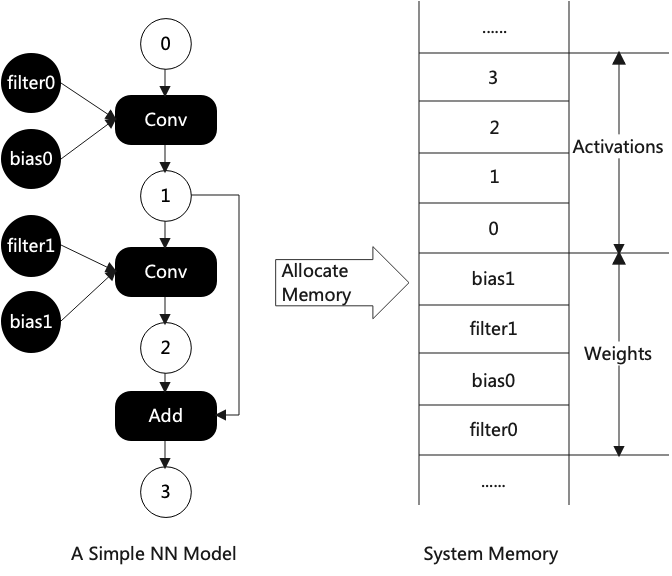}
   \caption{Buffer allocation in TPU-MLIR.}
   \label{fig:alloc}
\end{figure}

``inputs'' and ``outputs'' in ``InferenceParameter'' point to input buffers and output buffers of the operation. All buffers that tensor needed would be allocated after mlir file  were loaded. Each buffer size is calculated from Value's type. For example, the  \tensor{1x32x100x100xf32} needs $1\times32\times100\times100\times{\sf sizeof(float)} = 1280000 {\sf bytes}$. Figure~\ref{fig:alloc} is an example.

Weights are allocated and loaded first, and then activations are allocated. Before inference, inputs of the model will be loaded to input buffers. And then, run inference. After inference, results are stored in each activation buffers.

``handle'' in ``InferenceParameter'' is used to point third-party excute engine, and it is optional.

``InferenceInterface'' has three functions: ``init'', ``inference'', ``deinit''. ``init'' and ``deinit'' are used to init and deinit handle of third-party engine if needed, or do nothing. ``inference'' is used to run inference with ``inputs'' in ``InferenceParameter'' and store results in ``outputs'' of ``InferenceParameter''.

\subsection{Layer Group}

\begin{figure}[t]
   \centering
   \includegraphics[scale=0.4]{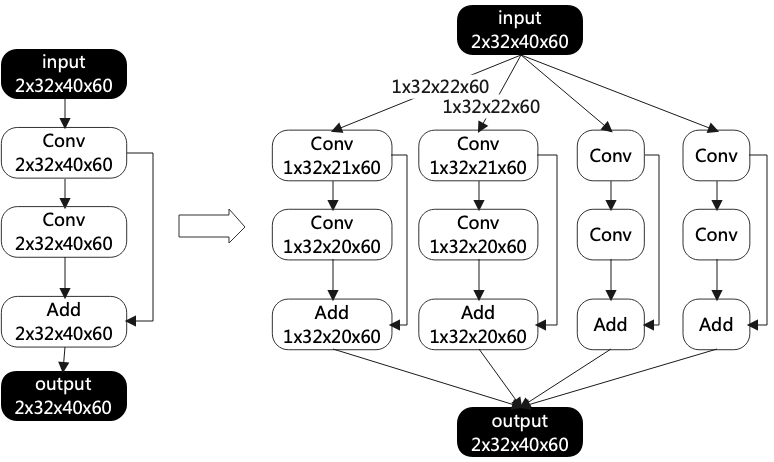}
   \caption{Slice the H dimensional in a layer group.}
   \label{fig:lg}
\end{figure}

Layer group in TPU-MLIR means some layers composed into one group execute in the TPU chip. The layer here is the same thing as the operation in MLIR. Typically, RAM on a chip is tiny, such as 256KB, while DDR off-chip is very large, such as 4GB. We need layers to run on the chip successively to achieve high performance, but the RAM on a chip is too small to support it. So we slice layers into small pieces to make sure layers in a group can run successively. Usually, we slice layers by N or H dimension. Figure~\ref{fig:lg} shows an example.

In mlir, we define group attributes for tpu operations:

\begin{lstlisting}[style=tblgen-code]
def Tpu_LayerGroupAttr : Tpu_Attr<"LayerGroup", "lg"> {
  let summary = "Structure of layer group parameters";
  let parameters = (ins
    "int64_t":$out_addr,
    "int64_t":$out_size,
    "int64_t":$buffer_addr,
    "int64_t":$buffer_size,
    "bool":$eu_align,
    ArrayRefParameter<"int64_t">:$h_idx,
    ArrayRefParameter<"int64_t">:$h_slice,
    ArrayRefParameter<"int64_t">:$n_idx,
    ArrayRefParameter<"int64_t">:$n_slice
  );
  let assemblyFormat = "`<` struct(params) `>`";
}
\end{lstlisting}

Different architecture TPU may have different attributes, attributes in ``LayerGroup'' are examples:

\begingroup
\renewcommand\labelenumi{\theenumi)}
\begin{enumerate}
\item out\_addr: output address in RAM on chip
\item out\_size: output memory size in RAM on chip
\item buffer\_addr: buffer address for operation in RAM on chip
\item buffer\_size: buffer size in RAM on chip
\item eu\_align: whether data arranged in RAM on chip is aligned
\item h\_idx: offset positions in h dimension as h has been sliced
\item h\_slice: size of each piece after sliced
\item n\_idx, n\_slice: for n dimension slice
\end{enumerate}
\endgroup

\begin{lstlisting}[style=mlir-code, caption={MLIR file with layer groups.}, label={lst:layer-group}, frame=lines, float=*t]
    %0 = "top.None"() : ()
    %1 = "top.Weight"() : ()
    %2 = "tpu.Group"(%arg0) ({
      %3 = "tpu.Load"(%arg0) {group_info = #tpu.lg<...>}
      %4 = "tpu.Cast"(%3) {group_info = #tpu.lg<...>}
      %5 = "tpu.Load"(%1) {group_info = #tpu.lg<...>}
      %6 = "tpu.Conv2D"(%4, %5, %0) {group_info = #tpu.lg<...>}
      %7 = "tpu.Cast"(%6) {group_info = #tpu.lg<...>}
      %8 = "tpu.Store"(%7) {group_info = #tpu.lg<...>}
      tpu.Yield %8 : tensor<1x32x100x100xf32, 4295618560 : i64>
    }) {hsecs = 1 : i64, nsecs = 1 : i64}
    return %2 : tensor<1x32x100x100xf32, 4295618560 : i64>
\end{lstlisting}

MLIR file with groups, like this Listing~\ref{lst:layer-group} (to make it simple, we have removed unrelated info from the file):
Layers in a group will execute on a chip successively, and the DMA will load data from DDR off-chip to RAM on-chip and store results back to DDR at the frontier of each group.

\subsection{Workflow}

This section we discuss the workflow of TPU-MLIR, expecially the main passes.

\begingroup
\renewcommand\labelenumi{\theenumi)}
\begin{enumerate}
\item OnnxConverter: use python interface to convert ONNX NN models to the TOP dialect mlir.

\item Canolicalize for TOP: do graph optimization on top operations.
  For example, we fuse top::ReluOp into top::ConvOp,  and we use depthwise conv to take the place of the batchNorn operation.

\item Calibration for TOP: use a few sampled inputs to do inference by top mlir file, and get every intermediate result, to stat proper min/max threshold.
  We use quant::CalibratedQuantizedType to express these calibration informations.
  For example, a value type is \tensor{1x16x100x100xf32}, and it's calibration informations are: min = -4.178, max = 4.493, threshold = 4.30.
  Then new type would be \tensor{1x16x100x100x!quant.calibrated$\langle$f32$\langle$-4.178:4.493$\rangle\rangle$} for
  asymmetric quantizaion, and \tensor{1x16x100x100x!quant.calibrated$\langle$f32$\langle$-4.30:4.30$\rangle\rangle$} for symmetric quantization.
  Do calibation only for int8 quantizaiton, and there is no need to do it for float convertion.

\item Conversion for TOP: convert top operations to tpu operations. We have discussed it above.

\item Layer group for TPU: determine groups of operations to execute successively in ram on tpu. We have discussed it above.

\item Memory assign for TPU: after TPU operations are ready, all operations out of group need to assign memory in DDR, especially assign physical address. We set physical address in tensor type, such as 4295618560 in \tensor{1x32x100x100xf32, 4295618560:i64}. We don't discuss how to assign memory by an optimal solution here.

\item Codegen for TPU: each TPU operation has codegen interface for different chips and has a corresponding TPU commands packaged in one kernel API. So what codegen to do is here just to call these APIs for each tpu operations, and collect commands to store in one model.
\end{enumerate}
\endgroup


\section{Conclusion}\label{sec:conclusion}
We are developing TPU-MLIR to compile NN models for TPU.
We design the TOP and TPU dialects as device-independent and device-dependent, respectively.
We convert NN models to Top dialect as device independent and convert TOP to TPU for various chips and types.
We define WeightOp for weight operation and store weight data in the NumPy npz file.
We design 'InferenceInterface' for top and tpu to ensure correct conversions.
In the future, we will try to support more TPU chips and NN models with various NN frameworks.

\bibliographystyle{ieee}
\bibliography{bibliography}

\end{document}